# A METHODOLOGY TO IDENTIFY THE LEVEL OF REUSE USING TEMPLATE FACTORS

N Md Jubair Basha[1] and Dr Chandra Mohan[2]

[1] Assistant Professor, IT Department, Muffakham Jah College of Engineering & Technology
Hyderabad, India.
`jubairbasha@mjcollege.ac.in`

[2] Associate Professor, CSE Department, JNT Univerity College of Engineering,
Hyderabad, India.
`c_miryala@yahoo.com`

## ABSTRACT

*To build large scale software systems, Component Based Software Engineering (CBSE) has played a vital role. The current practices of software industry demands more development of a software within time and budget which is highly productive to them. It became so necessary to achieve how effectively the software component is reusable. In order to meet this, the component level reuse, in terms of both class and method level can be possibly done. The traditional approaches are presented in the literature upto the level of extent of achievement of reuse. Any how still effective reuse is a challenging issue as a part. In this paper, a methodology has proposed for the identification of reuse level which has been considered by the using reuse metrics such as the Class Template Factor(CTF) and Method Template Factor(MTF). By considering these measures makes easy to identify the level of reuse so that helps in the growth the productivity in the organization.*

## 1. INTRODUCTION

In order to build the large scale software systems, Component Based Software Engineering has played a vital role. Reduction of Cost [1] and shorter development i.e. within time gives a good prospect for increasing the productivity in the organization. Components are connected by assembling, adapting and wiring into a complete application [2]. Although there is no IEEE/ISO standard definition that we know of, one of the leading exponents in this area, Szyperski [2], defines a software component as follows:
"A software component is a unit of composition with contractually specified interfaces and explicit context dependencies only. A software component can be deployed independently and is subject to composition by third parties".

The development of quality product within time and budget can be achieved through the effective software reuse. With this the effort and time required to be spend for testing and maintenance of the software products will also decreases.

Different methods and tools are proposed in the literature which provides only subset of operation requirements of effective software reuse. Though some of the approaches needs identify the metrics for the components[21] both at class level and methods level. In this paper, a methodology for identification of reuse level has been considered by the approach of reuse metrics considering the Class Template Factor(CTF) and Method Template Factor(MTF). The remaining part of this paper is organized as follows: section-2 presents the advantages of reusing software systems, section-3 describes about the domain engineering with its process and approaches for the domain engineering, section-4 describes about the reuse metrics for identifying reuse level and the implementation of those metrics on the HR Portal System These metrics measures about the reuse level in both class and method wise. and section -5 concludes the paper.

## 2. SOFTWARE REUSE

Software Reuse is the use of available software i.e. legacy software or to build a new software from software knowledge which is already existing. Reusable assets might be any one of the both i.e. legacy software or software knowledge. Reusability is a property of a software asset that indicates it's probability of reuse [3]. Software Reuse means the process that use "designed software for reuse" again and again for many times [4]. The advantage of reusing a software is to manage the complexity of software development, improvement in product quality and makes faster production in the organization.

Recently, design reuse has become popular with (object-oriented) class libraries, application frameworks, design patterns and along with the source code [5]. Jianli et al. proposed two complementary methods for reusing existing components. Among them one allows component evolution itself, which is achieved with binary class level inheritance across component modules. The other is by defined semantic entity so that they can be assembled at compile time or bind at runtime. Although component containment still is the main reuse model that leads to contribute the software product lines development [6]. Regarding the components much information has to be grouped, maintained and processed for the extraction of the components. N Md Jubair Basha et al. [7] has proposed a strategy to identify the components using clustering approach for component reusability. After identification of the components, the proposed work leads towards the identification of level of reuse which has been occurring in the components of an application.

Software Reuse can be broadly divided into two categories viz. Product reuse and Process reuse. The product reuse contains the reuse of a software component and by producing a new component which results in the module integration and construction. The process reuse provides the reuse of legacy components from repository. These components may be either directly reused or may need a minor modification by depending it's requirements for the retrieval. The modified software component can be grouped or stored by versioning these components. The components may be classified and selected depending upon the required domain [8].

The building of a component is the basis for their use in many applications. Reuse does not meant as a side effect. Specification, construction and testing must all be done for reuse. This makes a component more expensive (up to 10 times) to develop a new software.

Several different criteria for a good component have been suggested. These criteria can be summarized in the following[2]:

- The component should represent an abstract manner. It should have high cohesion and offer only the operations needed to make it useful in an efficient manner. It must also have well defined interfaces, both syntactically and semantically. If two

- operations in two different components have the same name, they should act in a similar manner. But their style should be similar to facilitate **understanding.**
- The component must be **independent** of surrounding entities; it should be loosely connected and thus have low coupling to other units. An object-oriented philosophy leads to independence.
- The component should be **general** abstraction which is useful in several applications without having to go unnecessary changes.

Understandability must be internal as well as external. Since good components will have a long life, they will be maintained for a long time.

The component system includes the selecting, classifying and managing or organizing the components which are available in the repository and also adds the developed new components in the repository. The component or work product repository should be spread throughout the development organization and that the components are accessible. The component repository should preferably be shared between several different work products. So, it means that the component system should serve on multiple projects. Whenever the new projects to be taken up, then the relevant components shall be needed for the development process. The project proposals should be reviewed by a group consisting of experienced designers and also someone from component department forming a software component committee. They should judge whether the proposed components are needs to be developed or not. If it is decided to construct the component, it is forwarded to component construction with a deadline. When ready, it is added to the component repository which then takes a new version state as showed in the Figure 1. As component is being used, the software component group should analyze it's value. Which component is used most? Which are not used at all? How much you gain from the components? This analysis helps to build the component system.

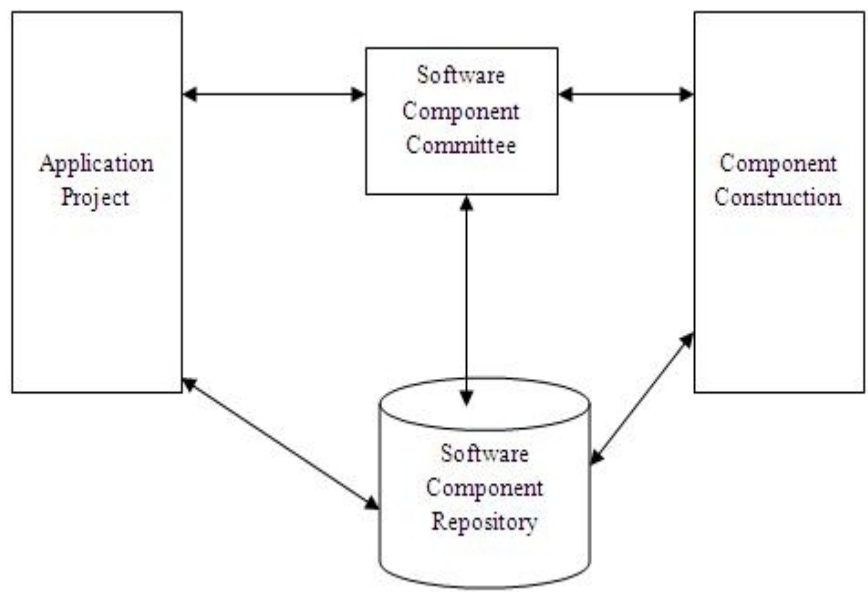

Figure: 1. An Organization for Component Management

## 3. DOMAIN ENGINEERING

The method of identifying the objects and operations for a class of similar systems helps in improving the software reuse. i.e. for a particular domain. In the terms of software engineering, domains are application areas [9].

There are many definitions about a *domain*. Czarnecki's defines [10]:" an area of knowledge scoped to maximize the satisfaction of the requirements of stakeholders, which includes concepts and terminology understood by practitioners in the area and the knowledge of how to build (part of) systems in the area".
Domain Engineering can be treated as a process where the reusable components are builded and organized and in which the architecture maps the requirements of the domain which has designed based on the domain[11].

Domain Engineering can be stated by the identification of the candidate domains and performing domain analysis and domain implementation which includes both application engineering and component engineering. Domain Analysis is a rigorous process of creating and maintaining the reuse infrastructure in a particular domain. The main purpose of domain analysis is to make the whole work product or component information easily and readily available. The relevant components (if available) has to be extracted from the repository rather than building the new components from the scratch for a particular domain[9].

Domain Analysis mainly focuses on reusability of analysis and design, but not code. This can be achieved by building common architectures, generic models or specialized languages that additionally improve the software development process in the specific problem area of the domain. A vertical domain is a specific class of systems. A horizontal domain contains general software parts being used across multiple vertical domains. Mathematical methods libraries container classes and UNIX tools are the examples of horizontal reuse. The purpose of domain engineering is to identify objects and operations of a class in a particular problem domain [9].

In the process of domain analysis, each component identified can be categorized as follows.

- General-purpose components : These components can be used in various applications of different domains (horizontal reuse).
- Domain-specific components :They are more specific and can be used in various applications of one domain (vertical reuse).
- Product-specific components : They are very specific and custom-built for a certain application, they are not reusable or only useful to a small extent.

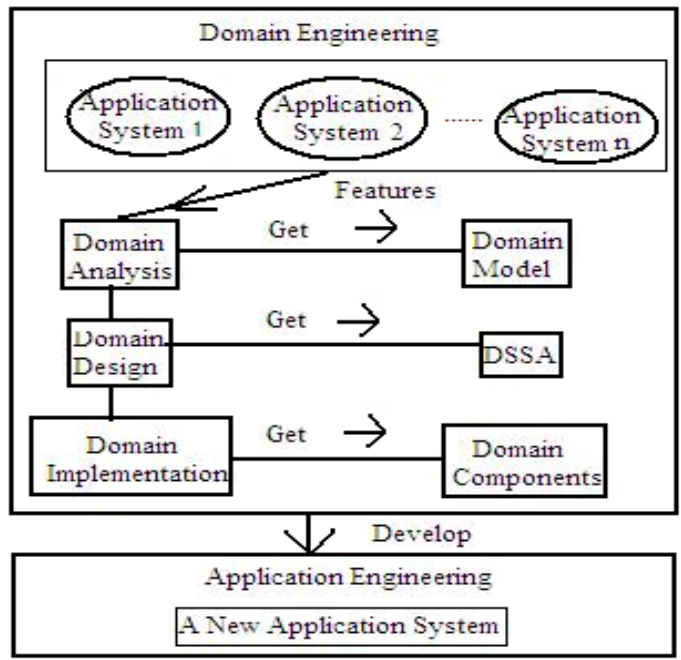

Figure:2. Domain Engineering Process

Domain engineering process [19] is depicted in figure 2. DE consists of three main stages i.e. domain analysis, domain design and domain implementation. For Domain Analysis support, DARE-COTS tool is presented [3]. Initially, in a particular domain it is mandatory to get the universal and variable characteristics of group systems. By abstracting the characteristics, domain analysis model can be generated. Based on this model the domain specific software architecture can be designed and then reusable components will be generated and organized.

## 3.1. APPROACHES FOR DOMAIN ENGINEERING

There are several known domain engineering tools. Each of these tools specifies a subset of operation requirements[18]

- Domain Analysis and Reuse Environment (DARE) is a tool developed in 1998 to support capturing information from experts, documents and code. Captured domain information is stored in a database that typically contains a generic architecture for the domain and domain-specific reusable components. DARE provides a library search facility with a windowed interface to retrieve the stored domain information [12].
- Family-Oriented Abstraction, Specification and Translation (FAST) is a system family generating method based on an application modelling language(AML) and was guiding developers to create the tools needed to generate software product line using domain engineering phase and application engineering phase.
- Feature Oriented Reuse Method (FORM) as an extension to the Feature Oriented Domain Analysis (FODA), a systematic method of capturing and analyzing commonalities and differences of applications in a domain (features). By using the results to develop domain architectures and components and modelling to discover and understand and capture commonalities' and variability's of a product line [13].

International Journal of Software Engineering & Applications (IJSEA), Vol.3, No.5, September 2012

- Kobra (KomponentenbasierteAnwendungsentwicklung) is used for component-based development [3]. Kobra method consists of product line development, component based software development and frameworks to provide systematic approach to developing high quality component based application frameworks [14]. Kobra is "technology independent" in the sense that it can be used with all the three major component implementation technologies CORBA, Java Beans and COM.
- Product Line UML-Based Software Engineering (PLUS) is a model-driven evolutionary development approach for software product lines. Apart from the analyzing and modelling a single system, it provides a set of concepts and techniques to explicitly model the commonality and variability in a software product line. With these techniques, object oriented requirements, analysis and design models of software product lines are developed using UML 2.0[15].
- Component Oriented Reverse Engineering (CORE) is a systematic and concrete model used to identify and develop reusable software components by using the reverse engineering techniques. This is used to extract architectural information and services from legacy system and later on convert the services into components [16].

## 4. REUSE METRICS

Software Design and Code Reuse will be made use in the object-oriented software development. The easiest way of reuse is being the use of a library class(of code), which matches the suits the requirements. Yap and Henderson-Sellers[22] presents two measures designed to evaluate the level of reuse of possible within classes.
The reuse ratio(U) is given by

$$U = \frac{\text{Number of Classes}}{\text{Total Number of Classes}}$$

For measuring reuse by using generic programming in the form of templates[23] has been proposed a set of metrics. The metric Method Template Factor (MTF) is defined as the ratio of the number of methods using method templates to the total number of methods as shown below.

$$MTF = \frac{\text{Number of methods using method templates}}{\text{Total Number of methods}}$$

Consider a system, with methods M1,....,Mn. Then,

$$MTM = \frac{\text{uses\_MT}(M_i)}{M_i}$$

where, uses_MT(Fi) = 1, iff method uses template method

0, otherwise

The following like wise code can be used to implement the MTM Metric

```
void function1() {
............}
template<class T>
void function2(T &x, T&y) {
...........}
void function3() {
............}
```

Figure: 3 A part of block Code for calculating the metric MTM

In figure 3, the value of metric MTM = 1/3.

The metric class template factor (CTF) is defined as the ratio of the number of classes using class template to the total number of classes as shown below:

$$CTF = \frac{\text{Number of classes using class templates}}{\text{Total Number of classes}}$$

Consider a system, with classes $C_1,........C_n$. Then

$$CTF = \frac{uses\_CT(C_i)}{C_i}$$

where, $uses\_CT(C_i) = $ 1, iff classes uses class template

0, otherwise

```
class A {
..........};
template<class T, int size>
class B{
T arr[size];
........};
```

Figure: 4 A part Source Code for calculating the metric CTF

In figure 4, the value of metric CTF = 1/2.

### 4.1 IDENTIFYINY REUSE LEVEL USING REUSE METRICS ON HR PORTAL SYSTEM

The HR Portal system has designed in such a way that the client can interact with the web tier and business tier and can connect to the Data Access Object(DAO) component. The web-tier component consists of the JSP's and Servlets. The Business tier consists of the EJB's. The DAO's consists of the classes with its objects communicating to the database.The web-tier components are HttpServlet, HRProcessServlet, Login Servlet, InterviewResultServlet and RegistrationServlet classes.The Business-tier components are EmployeeBean, InterviewResultsBean, HRProcessBean are the three stateless bean classes.The DAO components are BaseDAO, EmployeeDAO, InterviewDAO, HRDAO, ProcessDAO classes.

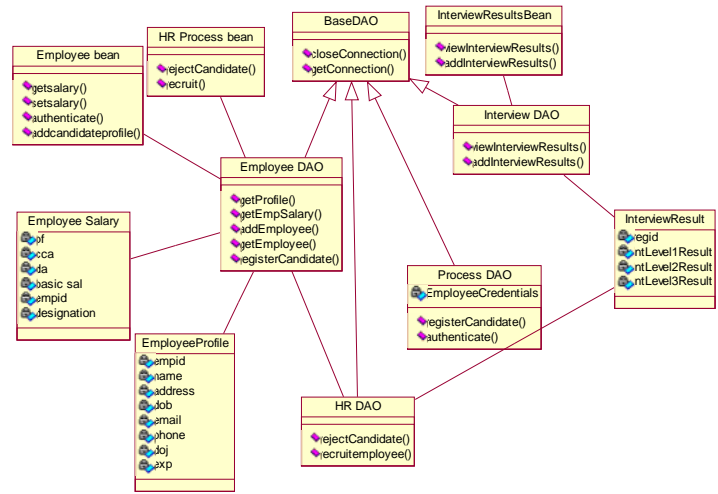

Figure 5:Business-Tier Class Diagram for HR Portal

The design of Figure 5 is about the business tier class diagram in which the relevant classes are showed with its methods. The classes are EmployeeBean, HRProcess, BaseDAO, EmployeeDAO, InterviewDAO, HRDAO, ProcessDAO, EmployeeProfile, EmpSalary, InterviewResult, InterviewResultsBean. Among these classes there are three stateless beans and five are the data access objects. EmployeeBean, InterviewResultsBean, HRProcessBean are the three stateless bean classes.

The metric Class Template Factor (CTF) is defined as the ratio of the number of classes using class template to the total number of classes as shown below:

$$CTF = \frac{\text{Number of classes using class templates}}{\text{Total Number of classes}}$$

Consider a system, with classes C1,........Cn. Then

$$CTF = \frac{uses\_CT(C_i)}{C_i}$$

where, $uses\_CT(C_i) = $ 1, iff classes uses class template

0, otherwise

From the above said metrics, CTF can be calculated on to the HR Portal system,

$$CTF = \frac{1+1+1}{11}$$

Under the EmployeeBean class, the getSalary(), authenticate(), addCandidatePofile() methods are designed. The getSalary() will give the salary of the employee which is reflected from the EmpSalary class and further connected to the EmployeeDAO object of the database. The authenticate() method will give registered user can sign in the Login page whose user mane are available in the database. The addCandidateProfile() is designed when the new registered user added to the BaseDAO. InterviewRessultsBean class consists of methods viewInterviewResults() and addInterviewResults(). The viewInterviewResults() methods is designed to check the result of the candidate who is pass or fail. The addInterviewresults() method is designed to add the Interviewed Results of the candidate depends upon the InterviewLevels. HRProcessBean class consists of methods rejectCandidate() and recruit().Depending upon the InterviewResltsBean, the recruit() and rejectCandidate() methods are designed to display the result. Thus, all the stateless beans in the HR Portal are designed.

The BaseDAO, EmployeeDAO, InterviewDAO, HRDAO, ProcessDAO are designed as the data access objects to the database. Among these objects, BaseDAO is the root for the other objects such as InterviewDAO, HRDAO, EmployeeDAO, ProcessDAO. BaseDAO class contains getConnection() and closeConnection() methods. The purpose of designing the getConnection() method is to connect to the database and closeConnection() method is to close the database connections. The EmployeeDAO consists of methods as getProfile(), getEmpSalary(), addEmployee(), getEmployee(), registerCandidate(). The design of these methods is based on the employee details. The getProfile() method is designed to give the Employee details such as empid, name, address, date of birth, email, phone, date of joining, experience. The getEmpSalary() method is to give the employee salary as pf, cca, hra, da, basicSal, empId, designation. The addEmployee() method is designed to add the employee to the database. The getEmployee() method is designed to give the Employee details such as empid, name, address, date of birth, email, phone, date of joining, experience. The registerCandidate() method is designed to register the candidate with his registration details which were specified in RegistrationServlet.The database tables are Employee Profile, Employee Salary, Interview Result and Employee Credentials.

The metric Method Template Factor (MTM) is defined as the ratio of the number of methods using method templates to the total number of methods as shown below

$$MTM = \frac{\text{Number of methods using method templates}}{\text{Total Number of methods}}$$

Consider a system, with methods M1,....,Mn. Then,

$$MTM = \frac{\text{uses\_MT}(M_i)}{M_i}$$

where, uses_MT(Mi) = 1, iff method uses template method

0, otherwise

The Method Template Factor (MTM) of Employee Bean

$$MTM_{EMPLOYEEBEAN} = 1/3$$

The Method Template Factor (MTM) of HRProcess

$$MTM_{HRPROCESS} = 1/2$$

The Method Template Factor (MTM) of BaseDAO

$$MTM_{BASEDAO} = 2/3$$

The Method Template Factor (MTM) of InterviewDAO

$$MTM_{INTERVIEWDAO} = 1/2$$

The Method Template Factor (MTM) of HRDAO

$$MTM_{HRDAO} = 1/3$$

The Method Template Factor (MTM) of ProcessDAO

$$MTM_{PROCESSDAO} = 1/3$$

The Method Template Factor (MTF) of EmployeeDAO

$$MTF_{EMPLOYEEDAO} = 2/3$$

The Method Template Factor (MTF) of EmployeeProfile

$$MTF_{EMPLOYEEPROFILE} = 1/2$$

The Method Template Factor (MTF) of EmpSalary

$$MTF_{EMPSALARY} = 1/2$$

The Method Template Factor (MTF) of InterviewResult

$$MTF_{INTERVIEWRESULT} = 1/4$$

The Method Template Factor (MTF) of InterviewResultsBean

$$MTF_{INTERVIEWRESUTTBEAN} = 1/2$$

With these metrics, it is easy to identify the level of reuse in terms of classes and methods. The above metrics gives better understanding of the reuse levels in both class wise and method wise. This results in the extent of reuse has been occurred in the class and number of methods with their part of block code which has effectively reused.

## 5. CONCLUSIONS

As there is a need to identify the measure of reuse level of the components in terms of both class and method level. A methodology has been proposed for the identification of reuse level which has considered by the approach of reuse metrics and by implementing the Class Template Factor(CTF) and Method Template Factor(MTF). This criteria will helps in identifying the level of reusability of the components in an application. As a future scenario, it is needed to implement this methodology to realize on any domain with some results. By considering this methodology, the productivity can be grown easily in the organization.

### ACKNOWLEDGEMENTS

The work was partly supported by the R & D Cell of Muffakham Jah College of Engineering & Technology, Hyderabad, India. The authors would like to thank to all the people from Industry and Academia for their active support.

## Authors

**N Md Jubair Basha** received his B.Tech. (IT) and M.Tech (IT) from JNTUH, Hyderabad. He is presently working as Assistant Professor in Department of Information Technology, Muffakham Jah College of Engineering and Technology, Hyderabad, India. His research interest includes Software Reusability, Component Based Software Development, Data Mining and Cryptography. He has published many research papers in various National/International Conferences and Journals. He is an active member of IEEE and CSI. 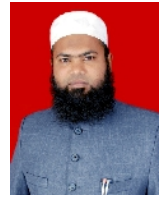

M. Chandra Mohan received B.E. (EEE) degree from Osmania University in 1994. He worked as Assistant Engineer in AP State Electricity Board (APSEB) for 7 years (1994-2001). He completed his M.Tech. (CS&E) from Osmania University in 2000. He is working in JNT University Hyderabad since 2001. Presently he is working as an Associate Professor in Dept of CS&E in JNTUH College of Engineering Hyderabad, JNT University Hyderabad. He is the recipient of 3 Gold Medals from Osmania University at the graduate level by securing University first rank. He completed his 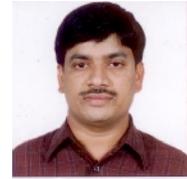
Ph.D in 2010 from JNTU Hyderabad in Computer Science & Engineering. He has published 13 research publications in various National and International Journals and conferences.